\documentclass[slac_one]{revtex4}
\usepackage{graphicx}
\usepackage{fancyhdr}
\begin{document}

\title{{\small{2005 ALCPG \& ILC Workshops - Snowmass,
U.S.A.}}\\ 
\vspace{12pt}
Branon Phenomenology} 

%

\author{J. A. R. Cembranos}
\affiliation{Department of Physics and Astronomy,
 University of California, Irvine, CA 92697 USA}

\author{A. Dobado}
\affiliation{Departamento de F\'{\i}sica Te\'orica I,
Universidad Complutense de Madrid, 28040 Madrid, Spain}

\author{A. L. Maroto}
\affiliation{Departamento de F\'{\i}sica Te\'orica I,
Universidad Complutense de Madrid, 28040 Madrid, Spain}

\begin{abstract}
In brane-world models with low tension, the fluctuations of the brane along the extra dimensions (branons) are the only relevant new low-energy modes. Such branon fields are in general massive, stable and weakly interacting, and accordingly they are natural candidates to explain the universe missing mass problem. On the other hand, the branon phenomenology is very rich because they couple to all the Standard Model particles. Distinctive branon signals could be observed in next colliders and in dark matter searches.
\end{abstract}

\maketitle
\thispagestyle{fancy}
\section{INTRODUCTION}
In flexible brane-world (BW) models, branons are the
only new relevant low-energy particles \cite{DoMa, GB}. The SM-branon
low-energy effective Lagrangian reads \cite{DoMa,BSky,ACDM}:
\begin{eqnarray}
{\mathcal L}_{Br}&=& \frac{1}{2}g^{\mu\nu}\partial_{\mu}\pi^\alpha
\partial_{\nu}\pi^\alpha-\frac{1}{2}M^2\pi^\alpha\pi^\alpha
\frac{1}{8f^4}(4\partial_{\mu}\pi^\alpha
\partial_{\nu}\pi^\alpha-M^2\pi^\alpha\pi^\alpha g_{\mu\nu})
T^{\mu\nu}\,.\label{lag}
\end{eqnarray}
Branons interact by pairs with the SM energy-momentum tensor $T^{\mu\nu}$,
and their couplings are suppressed by the brane tension $f^4$. In
fact, they are generically stable and weakly interacting. These
features make them natural dark matter \cite{CDM,M} candidates. On
the other hand,  the branon phenomenology is very rich since they
are coupled with the entire SM. The branon signals can be
characterized by their number $N$, the brane tension scale $f$,
and their masses $M$. For example, from the effective action given in
Equation (\ref{lag}), one can calculate the relevant cross-sections for
different branon searches in colliders (see Table \ref{tabHad} and References
\cite{ACDM,L3,CrSt}). 
\begin{table}
\centering \small{
\begin{tabular}{||c|cccc||}
\hline\hline Experiment

&
$\sqrt{s}$(TeV)& ${\mathcal
L}$(pb$^{-1}$)&$f_0$(GeV)&$M_0$(GeV)\\
\hline
HERA$^{\,1}
$& 0.3 & 110 &  16 & 152
\\
Tevatron-I$^{\,1}
$& 1.8 & 78 &   157 & 822
\\
Tevatron-I$^{\,2}
$ & 1.8 & 87 &  148 & 872
\\
LEP-II$^{\,2}
$& 0.2 & 600 &  180 & 103
\\
\hline
Tevatron-II$^{\,1}
$& 2.0 & $10^3$ &  256 & 902
\\
Tevatron-II$^{\,2}
$& 2.0 & $10^3$ &   240 & 952
\\
ILC$^{\,2}
$& 0.5 & $2\times 10^5$ &  400 & 250
\\
LHC$^{\,1}
$& 14 & $10^5$ &  1075 & 6481
\\
LHC$^{\,2}
$& 14 & $10^5$ &   797 & 6781
\\
CLIC$^{\,2}
$& 5 & $10^6$ &  2640 & 2500
\\
\hline\hline
\end{tabular}
} \caption{Summary of the main analysis related to
direct branon searches in collider experiments. All
 the results are
performed at the $95\;\%$ c.l. Two different channels have been
studied: the one marked with an upper index $^1\,$ is related to
monojet production, whereas the single photon is labelled with an
upper index $^2\,$. The table contains seven
experiments: HERA, LEP-II, the I and II Tevatron runs, ILC, LHC
and CLIC. The data corresponding to the four last
experiments are estimations, whereas the first three analysis have
been performed
 with real data.
$\sqrt{s}$ is the center of mass energy associated to the total
process; ${\mathcal L}$ is the total integrated luminosity;
 $f_0$, the bound in the brane tension scale for one
massless branon ($N=1$) and $M_0$ the limit on the branon mass for
small tension $f\rightarrow0$.}
\label{tabHad}
\end{table}

\section{RADIATIVE CORRECTIONS}

In addition to the corresponding missing energy signatures,
branons can also give rise to new effects through radiative
corrections. By integrating out the branon fields in the action
coming from ${\mathcal L}_{Br}$ it is possible to obtain an
effective action for the SM particles \cite{rad,radcorr} whose more
relevant terms are:
\begin{eqnarray}\label{eff}
{\mathcal L}_{SM}^{(1)}\simeq \frac{N \Lambda^4}{192(4\pi)^2f^8}
\left\{2T_{\mu\nu}T^{\mu\nu}+T_\mu^\mu T_\nu^\nu\right\}\,.
\end{eqnarray}

\begin{figure}[bt]
\begin{center}
\resizebox{7.5cm}{!} {\includegraphics{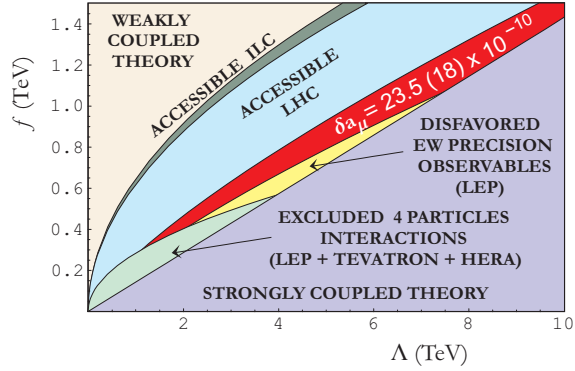}} \caption
{ Main limits from branon radiative corrections in
the $f-\Lambda$ plane for a model with $N=1$. The (red) central
area shows the region in which the branons account for the muon
magnetic moment deficit observed by the E821 Collaboration
\cite{BNL}, and at the same time, are consistent with present
collider experiments (whose main constraint comes from the Bhabha
scattering at LEP) and  electroweak precision observables.
Prospects for future colliders are also plotted.} \label{rad}
\end{center}
\end{figure}
%

\begin{table}[h]
\centering
\begin{tabular}{||c|c c c||} \hline
\hline
Experiment          & $\sqrt s$ (TeV) & ${\cal L}$ (pb$^{-1}$) &
$f^2/(N^{1/4}\Lambda)$ (GeV) \\ \hline
HERA$^{\,c}$        & 0.3             &  117                   & 52                           \\
Tevatron-I$^{\,a,\,b}$   & 1.8        &  127                   & 69                           \\
LEP-II$^{\,a}$      & 0.2             &  700                   & 59                           \\
LEP-II$^{\,b}$      & 0.2             &  700                   &
75                           \\ \hline
Tevatron-II$^{\,a,\,b}$  & 2.0        & $2 \times 10^3$        & 83                           \\ 
ILC$^{\,b}$         & 0.5             & $5\times 10^5$         & 261                          \\
ILC$^{\,b}$         & 1.0             & $2\times 10^5$         & 421                          \\ 
LHC$^{\,b}$     & 14              & $10^5$                 & 383
\\ \hline
\hline
\end{tabular}
\caption{\label{radcoll} 
Limits from virtual branon searches at colliders (results at the
$95\;\%$ c.l.) The indices $^{a,b,c}$ denote the two-photon,
$e^+e^-$ and $e^+p$ ($e^-p$) channels respectively. The first four
analysis have been performed with real data, whereas the final
four are estimations. The first two columns are the same as in
Table \ref{tabHad}, and the third one corresponds to the lower
bound on $f^2/(N^{1/4}\Lambda)$.}
\end{table}

\begin{figure}[bt]
\begin{center}
\resizebox{8cm}{!} {\includegraphics{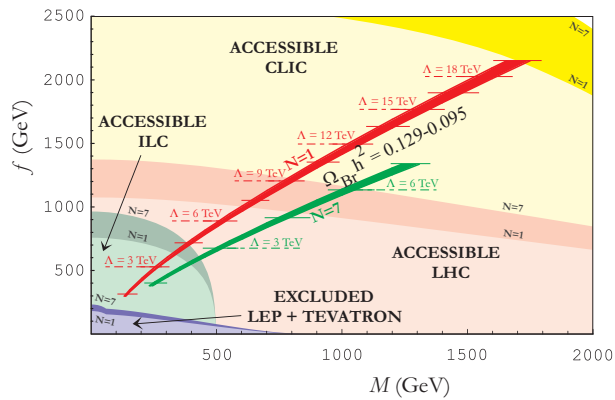}} \caption
{ Branon abundance in the range:
$\Omega_{Br}h^2=0.129 - 0.095$,
 in the $f-M$
plane (see \cite{CDM} for details). The regions are only plotted
for the preferred
values of the brane tension scale $f$. The central values of
$\Lambda$ from \cite{BNL} are also plotted. The lower area is
excluded by single-photon processes at LEP-II
and monojet signals at Tevatron-I \cite{ACDM}. The sensitivity of
future collider searches for real branon production are also
plotted (See \cite{ACDM} and Table \ref{tabHad}). The dependence
on the number of branons in the range $N=1-7$ can also be
observed.
} \label{CDM}
\end{center}
\end{figure}

$\Lambda$ being the cutoff setting the limit of
validity on the effective description of branon and SM dynamics
used here. This new parameter appears when dealing with branon
radiative corrections since the lagrangian in (\ref{lag}) is not
renormalizable. The most relevant contribution of branon radiative
correction at one loop is the modification of four-fermion
interactions and fermion pair annihilation into two gauge bosons
(see Table \ref{radcoll}). In addition , electroweak precision observables and the $\mu$
anomalous magnetic moment are corrected at two loops. In
\cite{radcorr} has been shown that branons can explain the
magnetic moment deficit of the muon found by the 821 Collaboration
at the Brookhaven Alternating Gradient Syncrotron \cite{BNL} and
be consistent with the rest of measurements (See Figure
\ref{rad}). But it is more remarkable that the same parameter
space which improve the fit of $g_\mu-2$ is able to explain the DM
content of the Universe. In fact, if the branon mass is of the
order of the electroweak scale, or more precisely, if the branon
mass is between $M\sim 100$ GeV and $M\sim 1.7$ TeV, branons could
form the total non baryonic DM abundance observed by WMAP
\cite{CDM} (see Figure \ref{CDM}).

\begin{acknowledgments}
JARC acknowledges the hospitality and collaboration of workshop organizers
and conveners, and economical support from the NSF and Fulbright OLP. 
This work is supported in part
by DGICYT (Spain) under project numbers
FPA 2004-02602 and FPA 2005-02327, by NSF grant No.~PHY--0239817
and by the Fulbright-MEC (Spain) program.
\end{acknowledgments}

\end{document}